\documentclass[doublecol]{epl2}
\usepackage{amsmath,bm}

\definecolor{darkgreen}{rgb}{0,0.6,0}
\definecolor{blue}{rgb}{0,0,0.8}
\definecolor{red}{rgb}{1.0,0.0,0.0}
\definecolor{lightblue}{rgb}{0.93,0.96,1}
\definecolor{darkblue}{rgb}{0.,0.,0.6}
\usepackage[colorlinks,linkcolor=darkgreen,citecolor=darkblue,urlcolor=blue]{hyperref}

\def\reff#1{(\ref{#1})}

\title{Synchronous versus asynchronous transport of a paramagnetic particle in a modulated ratchet potential}
\shorttitle{Transport of a paramagnetic particle in a modulated ratchet potential}

\author{Arthur V. Straube{$^1$\footnote{E-mails: \href{mailto:straube@physik.hu-berlin.de}{straube@physik.hu-berlin.de}, \href{mailto:ptierno@ub.edu}{ptierno@ub.edu};\newline
Published in \href{http://iopscience.iop.org/0295-5075/103/2/28001}{Europhys. Lett. (EPL) \textbf{103}, 28001 (2013)}}} \and Pietro Tierno$^{2,3}$}
\shortauthor{Arthur V. Straube and Pietro Tierno}

\institute{
  \inst{1} Department of Physics, Humboldt University of Berlin, Newtonstr. 15, D-12489 Berlin, Germany\\
  \inst{2} Departament de Estructura i Constituents de la Mat$\grave{e}$ria, Universitat de Barcelona, Av. Diagonal 647, 08028, Barcelona, Spain\\
  \inst{3} Institut de Nanoci$\grave{e}$ncia i Nanotecnologia IN$^2$UB, Universitat  de Barcelona, Barcelona, Spain\\
}

\abstract{We present a combined experimental and theoretical study describing the dynamical regimes displayed by a paramagnetic
colloidal particle externally driven above a stripe-patterned magnetic garnet film. A circularly polarized rotating magnetic field modulates the stray field of the garnet film and generates a translating periodic potential which induces particle motion. Increasing the driving frequency, we observe a transition from a phase-locked motion with constant speed to a sliding dynamics characterized by a lower speed due to the loss of synchronization with the traveling potential. We explain the experimental findings with an analytically tractable theoretical model and interpret the particle dynamics in the presence of thermal noise. The model is in good quantitative agreement with the experiments.}

\begin{document}

\maketitle

\section{Introduction}
Colloidal particles driven in a spatially periodic potential landscape
represent an ideal model system to study several fundamental phenomena, such as ratchet effects~\cite{ratchet},
depinning dynamics~\cite{pinning}, and phase synchronization~\cite{synchronization}. In particular, these systems are promising for applications concerned with the controlled transport and delivery of chemical or biological cargos
attached to functionalized particles~\cite{drug} or for high precision particle sorting and fractionation~\cite{sorting}.
Corrugated  potentials with periodicity on the colloidal scale can be
realized via different techniques, like by interfering laser beams~\cite{optical},
by using structured substrates characterized by high dielectric~\cite{electric} or
magnetic~\cite{magnetic} susceptibility contrast. In the latter case, uniaxial garnet films composed of
parallel stripes of oppositely magnetized ferromagnetic domains,
provide strong and externally tunable magnetic pinning sites, which can be used to control and
manipulate paramagnetic colloidal particles deposited above the films~\cite{garnet}.

In this context, a recent experimental work~\cite{Tie12} reports that a paramagnetic colloidal particle driven above a stripe-patterned
garnet film displays a series of dynamical phases, which can be controlled by varying the frequency of an externally applied
rotating magnetic field. The periodic potential characterized by a spatial period of few microns is generated by the stray field of a garnet film. The external magnetic field breaks the symmetry of the potential and creates a moving landscape which can transport the
particle. At low frequencies, the particle is synchronized with the external field and is therefore
transported with the speed of the traveling landscape. By increasing the driving frequency and beyond a critical
value, the particle desynchronizes with the translating potential, showing a complex sliding dynamics characterized
by a global decrease of its average speed.

In this letter, we study both experimentally and theoretically the dynamics of a micron-scale particle driven in such a
potential and focus on the loss of synchronization with the traveling landscape in the presence of thermal fluctuations.
In contrast to previous studies which provide rather complex expressions for the magnetic landscape and
neglect the effect of thermal noise, as, {e.g.}, in refs.~\cite{Tie12,Tie07,Verst},
we suggest a simple analytical expression for the potential allowing for a deep insight
into the transition from the completely synchronous to an asynchronous dynamic state.
Remarkably, despite its simplicity, the model not only captures the main physical features underlying the
transition but also demonstrates very good quantitative agreement with the measurements.
These findings can be crucial prerequisites for understanding collective effects in a similar system with interacting particles.

We start by describing the experimental system and the behaviour of a paramagnetic colloidal particle
at different driving frequencies. Then, we show how to derive an explicit expression
for the particle energy, and use a Langevin equation to describe the overdamped dynamics
of the particle in the moving landscape. Similarly to the experiments, the model predicts a frequency controlled transition
from a synchronous to an asynchronous dynamics, the latter being characterized by an alternation of running and oscillating states,
and a global decrease of the particle speed. By analyzing the average velocity and the distribution of the particle positions,
we explain the system behaviour as a loss of synchronization in presence of thermal noise.
\section{Experimental system}
\begin{figure}[!tb]
\begin{center}
\includegraphics[width=0.95\columnwidth,keepaspectratio]{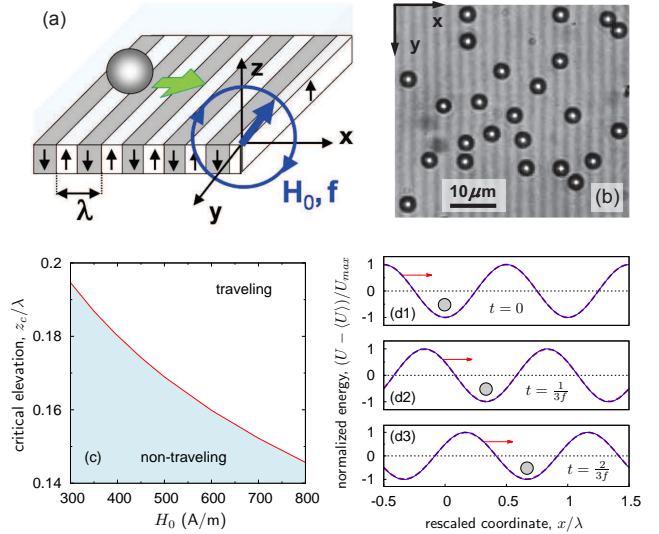}
\caption{(Color online) (a) Schematic of a single paramagnetic colloidal particle on top of a ferrite garnet film subjected to a rotating magnetic
field of the form of eq.~\reff{H-ext} with the frequency $f$ and amplitude $H_0$.
(b) Optical microscope image showing a diluted sample of particles (diameter $2.8 \, {\rm \mu m}$) and the magnetic pattern of the FGF (spatial periodicity $\lambda=2.6 \, {\rm \mu m}$).
(c) Critical elevation $z_{c}$ as a function of $H_0$, at which the energy potential $U$ starts to translate on average.
(d) Normalized energy potential, $(U-\langle U\rangle)/U_{max}$, of a single
particle as a function of $x$ at three different fractions of period, from top to bottom: $t=0$, $1/(3f)$, and $2/(3f)$.
The solid (blue) line shows the simplified potential as in
eq.~\reff{eq-U-simp}, and the bold dashed (red) line is for the general potential calculated via
eq.~\reff{eq-U} with the complex field given by eq.~\reff{H(w)} for $H_0/M_s=0.031$ and $z=0.923\lambda$. }
\label{fig:fig1}
\end{center}
\end{figure}

The periodic magnetic potential was generated by using a uniaxial ferrite
garnet film (FGF), grown by dipping liquid phase epitaxy~\cite{Tie09}.
The FGF was characterized by ferromagnetic domains having alternating magnetization with spatial period
$\lambda = 2.6 \, {\rm \mu m}$ and separated by Bloch walls (BWs),
i.e., narrow transition regions where the magnetic stray
field of the film is maximal, fig.~\ref{fig:fig1}(a).
We note that since the BWs exert attraction of the particle in the $z$ direction,
the particle elevation $z$ is essentially unaffected by thermal fluctuations and remains fixed,
resulting in an effectively two-dimensional system. To reduce the strong stray field by
about $40\%$ and to prevent from particle adhesion to the surface of the film, the FGF was coated with a layer of a
photoresist (AZ1512, Microchemicals) of thickness $l = 1 \, {\rm \mu m}$ \cite{Tie12}.

A water dispersion of monodisperse paramagnetic colloidal particles (Dynabeads M-270, Dynal)
of radius $a = 1.4 \, {\rm \mu m}$ and magnetic volume susceptibility $\chi=0.4$ was deposited above the FGF.
To induce particle motion, we applied an alternating ({AC}) magnetic field with the frequency $f$
and amplitude $H_0$ rotating
in the ($x,z$) plane and having circular polarization:
\begin{align}
& {\mathbf H}^{ext}(t) = H_0 \, [\cos (2\pi f t), \, 0, \, -\sin (2\pi f t)]\,. \label{H-ext}
\end{align}
Since the particle motion occurs only in a limited range of amplitudes, $200 \, {\rm A/m} \le H_0 \le 1000 \, {\rm A/m}$,\footnote{The particle motion was observed for amplitude of the applied field $200 \, {\rm A/m} < H_0 < 2000 \, {\rm A/m}$. For $H_0 < 200 \,{\rm A/m}$,
the field was not able to induce particle motion, while for $H_0 > 2000 \, {\rm A/m} $, the strong field distorts the BWs
and affects the particle trajectory.} in most of the experiments we fixed the amplitude $H_0=400 \, {\rm A/m}$ and varied the driving frequency $f$ in the range from $1 \, {\rm Hz}$ to $40 \, {\rm Hz}$. Figures \ref{fig:fig1}(a) and \ref{fig:fig1}(b) illustrate the experimental system, showing a schematic of a paramagnetic colloidal particle above the FGF and a polarization microscope image of a diluted suspension of particles above the magnetic stripes, respectively.
%
%
%
%

The external magnetic field of form \reff{H-ext} was applied by using two custom-made Helmholtz
coils arranged perpendicular to each other and with their main axis along the $x$ and $z$ directions.
The coils were mounted on the stage of a polarization microscope (eclipse NI-U, Nikon) equipped with a $100 \times
1.4$\,NA oil immersion objective and a CCD camera (Basler 601Fc) working at a temporal resolution of $60$ fps. AC currents flowing through the coils were generated via two independent current amplifiers (KEPCO BOP) controlled by an arbitrary waveform generator (TTi-TGA1244).
\section{Magnetic field above the FGF}
Let us consider a garnet film of thickness $d$ larger than its wavelength $\lambda$ and characterized by a
saturation magnetization $M_s$ subjected to an external rotating magnetic field of circular polarization, as
in eq.~\reff{H-ext}.
The global magnetic field ${\mathbf H}$ above the surface of the FGF,
$z \ge 0$, is given by the superposition
\begin{align}
& {\mathbf H}({\mathbf r},t)={\mathbf H}^{sub}({\mathbf r},t)+{\mathbf H}^{ext}(t)\,. \label{eq-Htot}
\end{align}
Here ${\mathbf H}^{sub}$ denotes the stray field of the substrate,
which obeys the equations $\nabla\times\mathbf{H}^{sub}=0$, $\nabla \cdot \mathbf{H}^{sub}=-M(x,t)\delta(z)$ supplemented by the periodicity condition, ${\mathbf H}^{sub}(x,z,t)={\mathbf H}^{sub}(x+\lambda,z,t)$, and being vanishing as $z \to \infty$. The choice $M(x,t)=\pm 2 M_s$ ensures that $H_z^{sub}(x,z,t)=\pm M_s$ at $z=0$.

The curl-free nature of ${\mathbf H}^{sub}$ allows us to work in terms of the ``electrostatic'' potential $\varphi$ introduced by $\mathbf{H}^{sub}=-\nabla\varphi$. Since there is no dependence of $y$, $\varphi$ satisfies the two-dimensional Laplace equation considered along with the condition of surface charge density $M(x,t)$ prescribed at $z=0$, which is solved by means of the conformal mapping technique. Accordingly, we introduce a complex potential $\Phi^{sub}(w)$ with $w=x+iz$ and $\varphi={\rm Re}[\Phi^{sub}]$. Because the fundamental solution for a point source at the origin is known to be $\Phi_0(w)=-1/(2\pi)\ln w$ \cite{Morse-Feshbach-book-53}, for our configuration of stripes of interchanging positive and negative surface charge $\pm 2M_s$, we have $\Phi^{sub}(w)=2M_s \sum_{n=-\infty}^{\infty}[\int_{x_n^{-}}^{x_n^{+}} dx \, \Phi_0(w-x) - \int_{x_n^{+}}^{x_{n+1}^{-}} dx \, \Phi_0(w-x)]$. Here, $x_n^{+}=n\lambda +\Delta(t)$ and $x_n^{-}=(n-1/2)\lambda -\Delta(t)$, $n=0, \pm 1, \pm 2, \dots$, are the positions of the BWs in the FGF and the quantity $\Delta(t)$ accounts for the fact that they can be moved upon application of an external field.

The field $\mathbf{H}$ is recovered from the potential via relation $\mathcal{H}(w)=-\partial_w \Phi(w)$, where $\mathcal{H}(w)=H_x-iH_z$ is the complex field with $H_x={\rm Re}\,\mathcal{H}(w)$ and $H_z=-{\rm Im}\,\mathcal{H}(w)$. Following eq.~\reff{eq-Htot}, the total field can be decomposed into the contributions caused by the substrate and external field, $\mathcal{H}=\mathcal{H}^{sub}+\mathcal{H}^{ext}$, where $\mathcal{H}^{ext}=H_x^{ext}-iH_z^{ext}$. For the substrate field, we find
$\mathcal{H}^{sub}(w)=-\partial_w \Phi^{sub}(w)=-(2M_s/\pi)\sum_{n=-\infty}^{\infty}[\ln(w-x_n^{+})-\ln(w-x_n^{-})]$.
We note that a similar approach is used in fluid mechanics to evaluate the velocity field
produced by streets of vortices considered as point sources of the same or
interchanging circulation \cite{Saffman-book-92}. Using the infinite product
representation, $\sin\pi z=\pi z\prod_{n=1}^{\infty}(1-z^2/n^2)$, the sums
over positive and negative sources can be evaluated explicitly to
yield $-(2M_s/\pi)\ln\sin[\pi(w-\Delta)/\lambda]$ and
$(2M_s/\pi)\ln\cos[\pi(w+\Delta)/\lambda]$, respectively.
As a result, the total field above the FGF can be represented as, cf. ref.~\cite{Verst}:
\begin{align}
\mathcal{H}(w) & =-\frac{2 M_s}{\pi}\ln \left(\frac{1-u_{-}}{1+u_{+}}\right)+H_x^{ext}-i H_z^{ext}\,, \label{H(w)}\\
u_{\pm}(w) & =\exp\left[\frac{i\pi}{2}\left(\frac{4w}{\lambda}\pm \frac{H_z^{ext}}{M_s}\right)\right]\,. \label{u(w)}
\end{align}
Since for this uniaxial FGF the displacement of the BWs is mainly caused by a
field perpendicular to the film, we put $\Delta(t)=\Delta_0 H_z^{ext}(t)/M_s$,
where $\Delta_0=\lambda/4$, as has been implicitly done previously~\cite{Tie07}.
In the absence of external field, eq.~\reff{H(w)} reduces to the result of Sonin~\cite{Sonin-02}.
It is also straightforward to show that
the field given by eq.~\reff{H(w)} can be alternatively represented via the complex potential
\begin{align}
\Phi(w) = \frac{i \lambda M_s}{\pi^2}\left[{\rm dilog}(1-u_{-})-{\rm dilog}(1+u_{+})\right]
-w \,\mathcal{H}^{ext} \nonumber 
\end{align}
with ${\rm dilog}(z)=\int_1^z \,d\zeta\,(1-\zeta)^{-1} \ln\zeta$, which is consistent with the expressions obtained in refs.~\cite{Tie12,Tie07}.

\section{Potential energy of a paramagnetic particle}
Under a magnetic field ${\mathbf H}({\mathbf r},t)$, a paramagnetic particle acquires a magnetic moment
${\mathbf m}=V \chi {\mathbf H}$ pointing along the field direction, where
$V= (4/3) \pi a^3$ is the particle volume.
The energy of interaction of a single magnetic dipole with the external field is
\begin{align}
U= -V\chi \mu_s \, {\mathbf H}^2= -V\chi \mu_s \left[({\rm Re}\,\mathcal{H})^2 + ({\rm Im}\,\mathcal{H})^2\right]\,, \label{eq-U}
\end{align}
where $\mu_s$ is the magnetic permeability of the medium.
We note that by using eq.~\reff{eq-U}, we assume that
the amplitude of external field is below the saturation magnetization, $H_0 \ll M_s$.

At a given elevation $z$, the local minima $x_{min}(t)$ of the energy landscape, eq.~\reff{eq-U}, remain either immobile on the average ($z<z_c$) or translate with a constant speed ($z>z_c$), implying different transport mechanisms. Generally, this speed averaged over one time period is given by $\overline{\dot x_{min}}=f\int_0^{1/f}\dot x_{min}(t)\,dt = v_m\theta(z-z_c)$, where $v_m =\lambda f$, $\theta$ is the Heaviside function, and the critical elevation $z_c$ depends on the amplitude $H_0$ of the field, see fig.~\ref{fig:fig1}(c). Recall that because of polymer coating, in our system the particle elevation $z \ge a+l\approx 2.4 \, {\rm \mu m} \approx \lambda = 2.6 \, {\rm \mu m}$. Therefore, for our range of field amplitudes $H_0$, we are always in the regime of the traveling potential ($z>z_c$).

We now show that in this case the potential energy, eq.~\reff{eq-U}, admits a very simple and accurate
approximation allowing for a straightforward interpretation of the dynamics of a single
particle. Since $H_z^{ext} \ll M_s$, we can write in the exponent of eq.~\reff{u(w)},
$u:=u_{\pm} \approx \exp(2\pi i x/\lambda) \exp(-2\pi z /\lambda)$.
Then, because $z\simeq \lambda$, $u$ can be treated as a small parameter to give $\ln[(1-u)/(1+u)]=-2u+\mathcal{O}(u^3)$. As a result, the field above the substrate is approximated by
\begin{align}
{\mathbf H}^{sub}({\mathbf r})= \frac{4 M_s} {\pi} \, {\rm e}^{-2\pi z/\lambda}\left( \cos \frac{2\pi x}{\lambda},\, 0,\,-\sin \frac{2\pi x}{\lambda}\right)\,. \nonumber
\end{align}
Evaluating eq.~\reff{eq-U} and omitting the terms independent of $x$, we 
arrive at the approximate potential describing the interaction of a particle with the field above the FGF
\begin{align}
\frac{U(x,t)}{U_0} = -\frac{8 H_0} {\pi M_s} \, {\rm e}^{-2\pi z/\lambda} \, \cos \left( \frac{2\pi x}{\lambda}-2\pi f t \right)\,, \label{eq-U-simp}
\end{align}
which is written down relative to the characteristic magnetic energy $U_0 = V\chi \mu_s M_s^2$.

In fig.~\ref{fig:fig1}(d), we compare the approximate expression for the potential given by eq.~\reff{eq-U-simp}
shown as a solid (blue) line and the general potential in eq. \reff{eq-U} evaluated via
eqs.~\reff{H(w)} and \reff{u(w)} shown as a bold dashed (red) line, which are in perfect agreement. Furthermore, fig.~\ref{fig:fig1}(d) shows that the energy potential presents a spatially periodic landscape characterized by minima at the positions $x_{min}(t)=n\lambda + v_m t$ ($n=0, \pm 1, \pm 2, \dots$) spaced by $\lambda$, which continuously translate with time with a constant speed $v_m = \lambda f$ along the $x$ axis. As a consequence, a particle sitting initially in one of such minima, can synchronize with the field and follow the moving landscape with the speed $v_m$ with a lag behind the minimum. This dynamic state is observed in experiments at relatively low frequencies. At high frequencies, however, the particle can lose its synchronization, which results in the decrease of its average speed.

\section{Theoretical model of particle motion}

We describe the overdamped dynamics of a paramagnetic particle subjected to
a deterministic potential $U(x,t)$ in the presence of thermal noise by using the following
Langevin equation~\cite{Haenggi-etal-05, Burada-etal-09}:
\begin{align}
\zeta\,\dot x(t) = -\frac{\partial U(x,t)}{\partial x}+\sqrt{2 \zeta k_B T} \, \xi(t)\,, \label{Langevin-dim}
\end{align}
Here, $\zeta=6\pi\eta a$ denotes the friction coefficient between the particle and the solvent
resulting from the Stokes drag, $\eta$ is the dynamic
viscosity of the medium (water), $k_B$ is the Boltzmann constant, and $T$
is the temperature. The stochastic force is modeled as a Gaussian white
noise obeying the properties: $\left<\xi(t)\right>=0$ and
$\left<\xi(t)\xi(t')\right>=\delta(t-t')$.

For convenience, we proceed to normalized variables by measuring
the length, time, magnetic field, and energy in the units of $\lambda$,
$\zeta\lambda^2/U_0$, $M_s$, $U_0$, respectively. As a result,
writing eq.~\reff{Langevin-dim} for the potential in eq.~\reff{eq-U-simp},
gives:
\begin{align}
\dot x(t) = -16 \,h_0 \, {\rm e}^{-2\pi z}\sin[2\pi (x -\tilde{f} t)] +\sqrt{2 \sigma} \, \xi(t)\,. \label{Langevin}
\end{align}
Here, we have introduced three dimensionless parameters,
related to the amplitude, $h_0= H_0/ M_s$, frequency, $\tilde{f}=(f \zeta \lambda^2)/U_0$,
of the external field, and to the intensity of thermal fluctuations, $\sigma=(k_B T)/U_0$.

In terms of a new variable, $q(t)=-x(t) +\tilde{f} t$, which makes the system autonomous by proceeding to the reference frame traveling with the potential, we obtain a stochastic Adler equation:
\begin{align}
\dot q(t) & = \tilde{f} - \tilde{f}_c \sin [2\pi q(t)] +\sqrt{2 \sigma} \, \xi(t)\, \label{Langevin}
\end{align}
with the ``critical'' frequency given by
\begin{align}
\tilde{f}_c =16 h_0 \,{\rm e}^{-2\pi z}\,. \label{Omega-c}
\end{align}

In the limiting case of negligible fluctuations, $\xi=0$,
the particle behaviour corresponds to the motion in a tilted ``washboard'' potential given by, $ \tilde{V}(q)=- \tilde{f}q - \tilde{f}_c\cos(2\pi q)/(2\pi)$. Note that $2\pi q$ plays the role of phase. By setting $\dot q =0$, we make sure that
depending on the frequency $\tilde{f}$, the system admits
two dynamical regimes separated by $\tilde{f}_c$. For $\tilde{f} < \tilde{f}_c$,
there exist local minima in the potential $\tilde{V}(q)$ and we have a stable solution,
$q_0=\arcsin(\tilde{f}/\tilde{f}_c)$, which corresponds to a ``phase-locked''
state in the co-moving reference frame. The motion of the particle is synchronized with the external field. The particle position in the laboratory reference frame is $x(t)=x_{min}(t)+\delta x$, where $\delta x=-q_0$, and hence, it generally lags behind a local minimum
of the potential. At low frequencies, $\tilde{f} \ll \tilde{f}_c$, the lag is small, $\delta x \propto -(\tilde{f}/\tilde{f}_c)$,
while at frequencies approaching $\tilde{f}_c$, it tends to its maximum absolute value,
$\delta x = -1/4$. Beyond the critical frequency, $\tilde{f} >\tilde{f}_c$, there are
no minima in $\tilde{V}(q)$. The particle cannot synchronize with
the field and starts to slide, which is typically referred to as the ``phase-drift'' or sliding regime.
Similar to synchronization problems~\cite{Pikovsky-etal-book-01, Adler-46},
we evaluate the time $\Delta t =\int_0^1 [\tilde{f}-\tilde{f_c}\sin(2\pi q)]^{-1}{\rm d}q =(\tilde{f}^2-\tilde{f}_c^2)^{-1/2}$ it takes the particle to travel the distance of one period $\Delta q=1$. As a result,
$\left<\dot q \right>=\Delta q/(\Delta t)$ and the
averaged speed $\left<\dot x\right>=\tilde{f}-\left<\dot q\right>$ normalized by its maximum value $v_m$ is given by:
\begin{align}
\frac{\left<\dot x\right>}{v_m} = \left\{
\begin{array}{ll}
1, & {\rm if} \;\; \tilde{f} < \tilde{f}_c\,, \\
1 - \sqrt{1-(\tilde{f}_c/\tilde{f})^2}\,, & {\rm if} \;\; \tilde{f} > \tilde{f}_c\,, \\
\end{array}
 \right. \label{speed-det}
\end{align}
which describes the transition from synchronous to asynchronous
dynamics encountered in several physical systems \cite{a-synchronous},
and is in agreement with the general interpretation of a traveling potential ratchet
(see Sec.~4.4 of ref.~\cite{Reimann-02}).

To address the situation involving thermal fluctuations,
we use a Fokker-Planck equation
corresponding to eq.~\reff{Langevin},
%
\begin{align}
\frac{\partial}{\partial t} P(q,t)=2\pi \sigma \frac{\partial}{\partial q}\left[\frac{dV(q)}{dq}+\frac{1}{2\pi}\frac{\partial}{\partial q}\right]P(q,t)\,, \label{FPE}
\end{align}
where $V(q)=\tilde{V}(q)/\sigma=-2\pi q D - D_c \cos(2\pi q)$, $D=\tilde{f}/(2\pi \sigma)$, and $D_c=\tilde{f}_c/(2\pi \sigma)$.
The stationary solution $P_0(q)$ of eq.~\reff{FPE} satisfying the periodicity requirement,
$P_0(q+1)=P_0(q)$, and the normalization condition, $\int_0^{1} P_0(q) \,{\rm d}q =1$,
is given by \cite{Stratonovich-book-67}:
\begin{align}
P_0(q)=\frac{1}{\mathcal{N}}\,\mathcal{P}(q), \;\;\; \mathcal{P}(q)={\rm e}^{-V(q)} \int_{q}^{q+1} {\rm e}^{V(q')} \,{\rm d}q'\,. \label{P0(y)}
\end{align}
Here, $\mathcal{N}={\rm e}^{-\pi D} \, |I_{iD}(D_c)|^2$ is
the normalization constant and $I_{i\nu}(x)$ is the modified Bessel
function of the first kind of an imaginary order.
Note that $P_0(\delta x)=P_0(-q)$ with $\delta x(t)=x(t)-v_m t$ describes the distribution of particle positions
relative to the deterministic trajectory with the constant speed $v_m=\tilde{f}$. Because of periodicity,
$P_0(\delta x)=P_0(x-x_{min})$ also corresponds to the distribution relative to the minima of the traveling potential.

Using the distribution in eq.~\reff{P0(y)}, we can average eq.~\reff{Langevin} to arrive at the Stratonovich formula ~\cite{Stratonovich-book-67, Risken-book}: $\left<\dot q\right>=\int_0^{1} \dot q \,P_0(q)\,{\rm d}q=\mathcal{N}^{-1}[1-{\rm e}^{V(q+1)-V(q)}]=2\sigma\,|I_{iD}(D_c)|^{-2}\sinh(\pi D)/(\pi D)$.
Transforming back to $x$, we end up with the generalization
of the relation given in eq.~\reff{speed-det} for the case of thermal fluctuations
\begin{align}
\frac{\left<\dot x\right>}{v_m} = 1-\frac{\sinh(\pi D)}{\pi D \,|I_{iD}(D_c)|^2}\,. \label{speed-stoch}
\end{align}
In the limit of $\sigma\to 0$, eq.~\reff{speed-stoch} reduces to its deterministic
counterpart, eq.~\reff{speed-det}.
%
\begin{figure}[!tb]
\begin{center}
\includegraphics[width=0.46\textwidth]{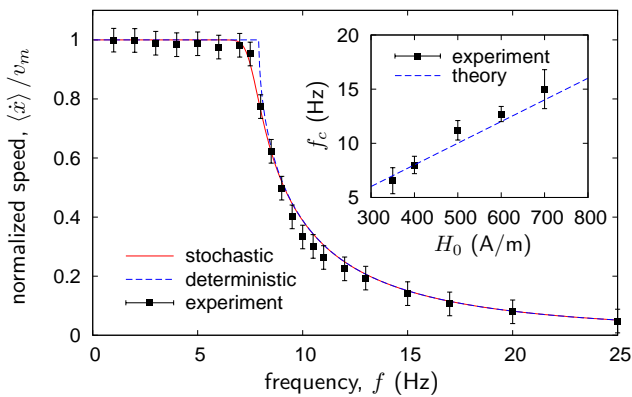}
\end{center}
\caption{(Color online)
Normalized particle velocity $\langle \dot x \rangle/v_m$ as a function of the driving frequency $f \, {\rm(Hz)}$ for a paramagnetic 
particle subjected to a rotating field with amplitude $H_0=400 \, {\rm A/m}$.
The experimental data (solid squares) are plotted together
with the theoretical lines. The dashed line shows the deterministic limit according to formula \reff{speed-det}, the solid line is for the general
case with thermal noise described by eq.~\reff{speed-stoch} with $\sigma=0.5 \times 10^{-5}$.
Inset shows the comparison of the critical frequency as a function of the field amplitude.
Markers represent experimental data, the dashed line
corresponds to eq.~\reff{Omega-c}.
}
\label{fig2}
\end{figure}
%

\section{Discussion and conclusions}

In order to validate the theoretical model, we perform a series of experiments
by measuring the average speed $\langle \dot{x} \rangle$ as
a function of frequency $f$ for a single paramagnetic particle driven by a rotating field above the FGF.
Figure~\ref{fig2} shows $\langle \dot{x} \rangle$ normalized by the maximum speed
for a fixed field amplitude $H_0=400 \,{\rm A/m}$.
The transition between the phase-locked regime and the asynchronous dynamics is observed at $f \approx 7.9 \, {\rm Hz}$.
In fig.~\ref{fig2} we fit the theoretical expressions for the deterministic, eq.~\reff{speed-det} (dashed line), and stochastic, eq.~\reff{speed-stoch} (solid line), averaged speed, against the experimental data
and find an almost perfect agreement by
accounting for
the thermal noise.
We fixed the dimensionless elevation $z=0.923$ (or, equivalently, $z=2.4 \; {\rm \mu m}$)
and varied the values of $h_0$ and $\sigma$ to achieve the best agreement between the data at $h_0=0.031$ and $\sigma=0.5 \times 10^{-5}$. The comparison between the theoretical fits using eq.~\reff{speed-det} (dashed line) and eq.~\reff{speed-stoch} (solid line) shows that the 
effect of thermal fluctuations is smoothing
the transition region between both regimes, slightly shifting the critical point.

In the inset of fig.~\ref{fig2} we show the critical frequency
extracted from the experimental data (squares) as a function of the field amplitude.
The theoretical prediction based on eq.~\reff{Omega-c} with the same value of $h_0$ as in fig.~\ref{fig2} is in agreement with the results of measurements.

In a previous work~\cite{Tie07} based on the use of FGF with large
ferromagnetic domains ($\lambda=10.9 \,{\rm \mu m}$),
it was possible to determine $M_s$ directly, i.e.,
by measuring the displacement of the BWs under an external
magnetic field normal to the film~\cite{Kooy60}.
For the small stripes used in this work, diffraction
limits the resolution capability of our polarization microscope,
making impossible the direct measurement of $M_s$. Therefore, using the experimental values for the particle elevation and amplitude of the magnetic field ($H_0= 400 \, {\rm A/m}$), we estimate the saturation magnetization of the FGF, $M_s =H_0/h_0\approx 1.3 \times 10^4 \, {\rm A/m}$.
Furthermore, our approach allows us to estimate the
strength of effective noise, which tells us whether
its origin is simply thermal or it could be due to other factors
like presence of disorder in the film or deformation of BWs~\cite{Tie10}.
%
%
\begin{figure}[!tb]
\begin{center}
\includegraphics[width=0.95\columnwidth,keepaspectratio]{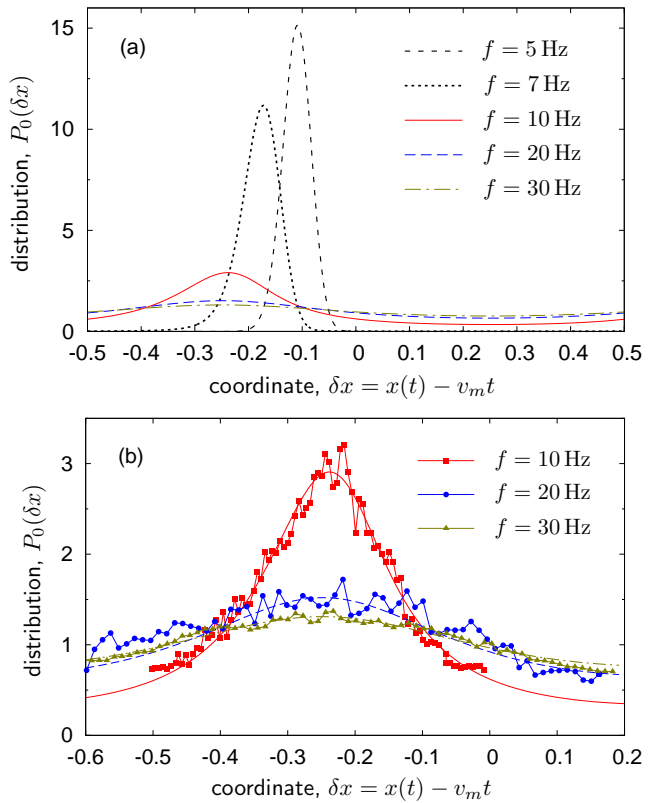}
\end{center}
\caption{(Color online)
Distribution of the particle positions, $P_0(\delta x)$, in the reference frame moving with the speed $v_m$ shown
for different driving frequencies $f$. (a) Theoretical predictions evaluated via eq.~\reff{P0(y)}.
(b) Comparison of the theoretical predictions with the experimental data (markers), smooth bold lines are theoretical
curves calculated via eq.~\reff{P0(y)} for dimensionless parameters
$h_0=0.031$, $z=0.923$, and $\sigma=0.5 \times 10^{-5}$.} \label{fig3}
\end{figure}
%
%
In particular, the obtained estimate $\sigma=0.5 \times 10^{-5}$ corresponds to an effective thermal energy
of $\approx 1.2 k_BT$, indicating that the driven particle is mainly
subjected to thermal fluctuations along the direction of motion.

To analyze the effect of fluctuations further, we extract the particle positions in the co-moving reference frame, $\delta x(t)=x(t)-v_m t$,
from several trajectories and calculate the corresponding probability density distribution $P_0(\delta x)=P_0(-q)$ for
different frequencies. Note that $\delta x$ also shows the shift of particles from the minimum of the potential at $\delta x=0$.
The results are shown in figs.~\ref{fig3}(a) and \ref{fig3}(b) together with the theoretical expressions calculated via eq.~\reff{P0(y)}. In agreement with the deterministic consideration, we see that in the phase-locked state at low frequencies the particle is localized near the minimum of the potential at $\delta x =0$. Note that the low-frequency limit of eq.~\reff{P0(y)} provides $P_0(\delta x)=\exp[D_c\cos (2 \pi\delta x)]/I_0(D_c)$, which is symmetric and centered around $\delta x=0$. As the frequency is increased, the particle starts to gradually lag behind the potential: the probability density distribution becomes wider and shifts towards negative values of $\delta x$. This tendency is also observed in the domain beyond the critical frequency, when the particle is in the asynchronous regime, as confirmed by experimental data, see fig.~\ref{fig3}(b).

In summary, we studied both experimentally and theoretically the dynamics of a paramagnetic colloidal particle magnetically driven above a stripe patterned garnet film.  We explored the dynamics in a wide range of frequencies and various field amplitudes, and observe, in all cases, a transition from a synchronous regime with a constant speed to an asynchronous one. In all regimes, with the growth in frequency the most probable position of a particle increasingly lags behind the minima of the moving landscape.
We put forward a simple and analytically tractable approximation for the potential that is in
quantitative agreement with the experimental findings. We show how thermal fluctuations affect the loss of synchronization and how the theoretical results can be matched with real measurements, where thermal noise is unavoidable. One the one hand, the suggested approach can be extended to other physical systems characterized by the similar generic transition, where the role of noise is neglected. On the other hand, this work not only provides a solid theoretical ground to recent experimental findings but it can be also used as a starting point for extensions towards inter-particle interactions, and thus collective effects, since it provides simple and accurate expressions for the magnetic potentials.

\acknowledgments
We acknowledge \Name{Tom H. Johansen} for the FGF,
and \Name{J. Ort\'in} for laboratory support.
\Name{AVS} was supported by the European Science Foundation (ESF),
project 5477 within the activity ``Exploring the Physics of Small Devices
(EPSD)''. \Name{PT} acknowledges support from the programs RYC-2011-07605,
FIS2011-13771-E, FIS2011-15948-E.

\end{document}